\documentclass{appolb}
\usepackage{graphicx}

\begin{document}
\title{Recent theoretical results for electromagnetically induced
ultraperipheral reactions of heavy ions%
\thanks{Presented at the XIII Workshop on Particle Correlations and
  Femtoscopy, 22-26 May 2018, Krak\'ow, Poland}%
}
\author{A. Szczurek
\address{Institute of Nuclear Physics, PAN\\
ul. Radzikowskiego 152, PL-31-342 Krak\'ow, Poland\\
also at Faculty of Mathematics and Natural Sciences,\\
University of Rzesz\'ow, Poland}
}
\maketitle
\begin{abstract}
We briefly review our works on ultraperipheral
heavy ion collisions. We discuss both $\gamma \gamma$
and rescattering of hadronic photon fluctuation induced by one nucleus 
in the collision partner. Production of one and two leptonic and
pionic and $p \bar p$ pairs is discussed as an example 
of photon-photon processes. The production of single vector mesons 
($\rho^0$ or $J/\psi$) is an example of the second category.
The double-scattering mechanisms of two $\rho^0$ meson production
is discussed in addition.
\end{abstract}

\section{Introduction}

The ultraperipheral collisions is a class of processes that were 
studied experimentally only recently at RHIC and the LHC.
The process can be viewed as a scattering of two clouds of photons
or a process of scattering of a photon (or photon hadronic fluctuations)
emitted by one nucleus on the second colliding nucleus.
In general, one is interested rather in processes with small particle
multiplicity which automatically means that the impact parameter
is greater than the sum of the radii of colliding nuclei.
Some of such processes were suggested long ago \cite{reviews}.
Only recently some experimental results were presented.
In the following we will present some results. In addition we will show
some other processes that could be also studied at the LHC.

\begin{figure}
\begin{center}
\includegraphics[width=5cm]{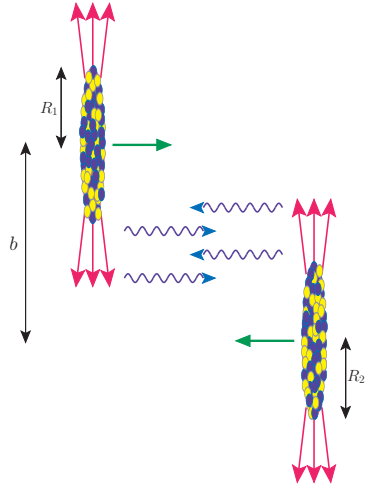}
\end{center}
\caption{A schematic view of the $\gamma \gamma$ induced processes.}
\label{fig:EPA}
\end{figure}

\begin{figure}
\begin{center}
\includegraphics[width=4cm]{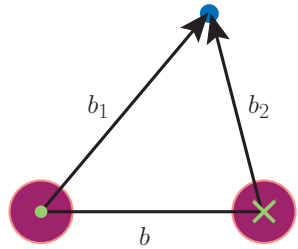}
\end{center}
\caption{The situation in the impact parameter space.}
\label{fig:impact_parameter}
\end{figure}

A schematic view of the photon induced processes is shown in 
Fig.\ref{fig:EPA} and the situation in the impact parameter space is
illustrated in Fig.\ref{fig:impact_parameter}.
When calculating the cross section in the equivalent photon
approximation in the impact parameter space ultraperipheral collisions
mean that the two circles, representing heavy ions, do not overlap
($b > R_A + R_B$) \cite{reviews}. 
It does not mean, however, that the processes of
photoproduction disapear in such a case. In this case they may also 
contribute and compete with other procesess characteristic for standard
($b < R_A + R_B$) heavy ion collisions. 
The situation/physics then strongly depends on the reaction.

Our detailed studies were presented in Refs.[2-16]. 
In this presentation we discuss different processes except
light-by-light processes that were discussed in \cite{Klusek_talk}.
Here we only sketch some selected results.

\section{A brief review of our results for UPC}

\begin{figure}[!h]
\begin{center}
\includegraphics[width=4.5cm]{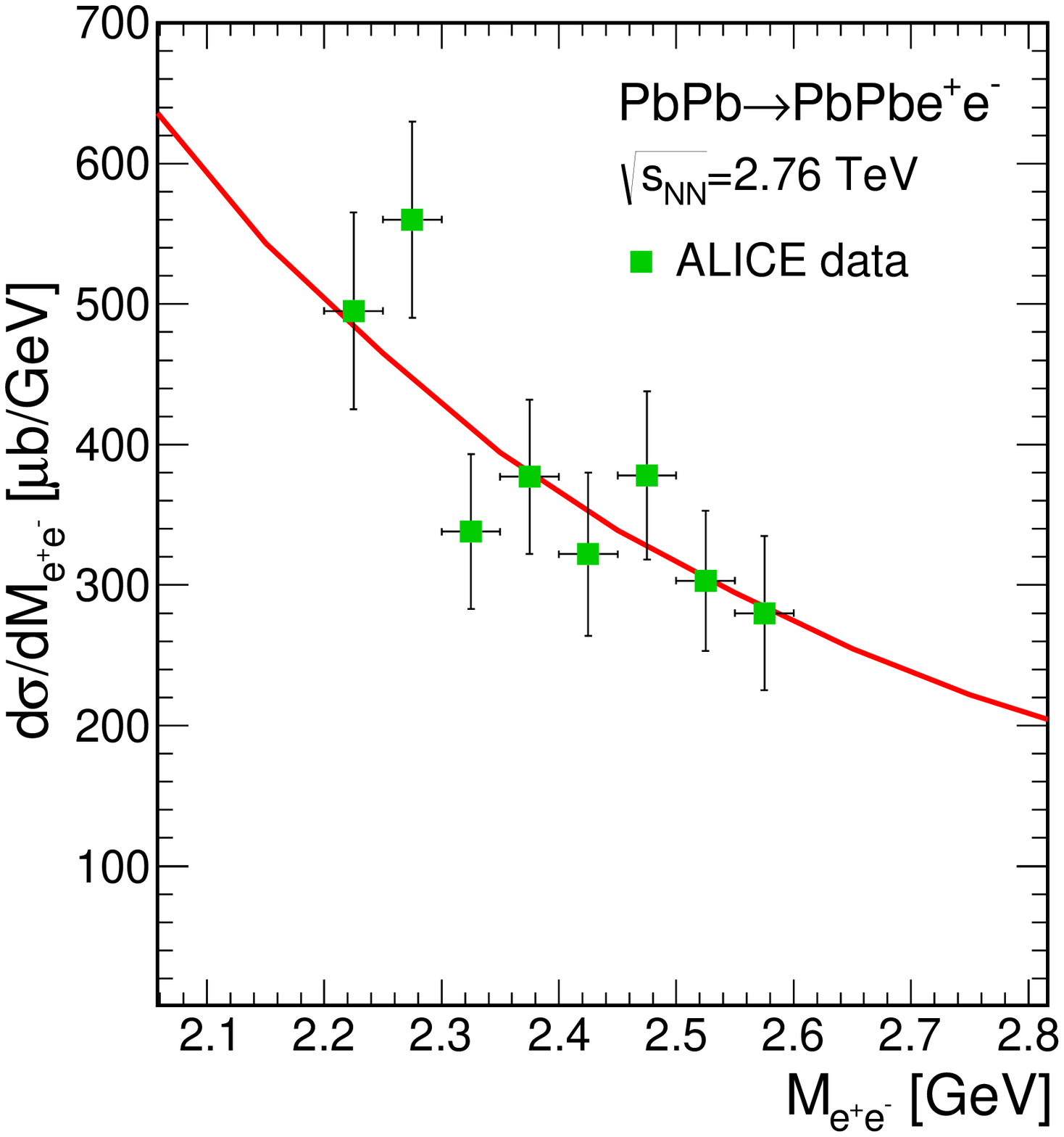}
\includegraphics[width=4.5cm]{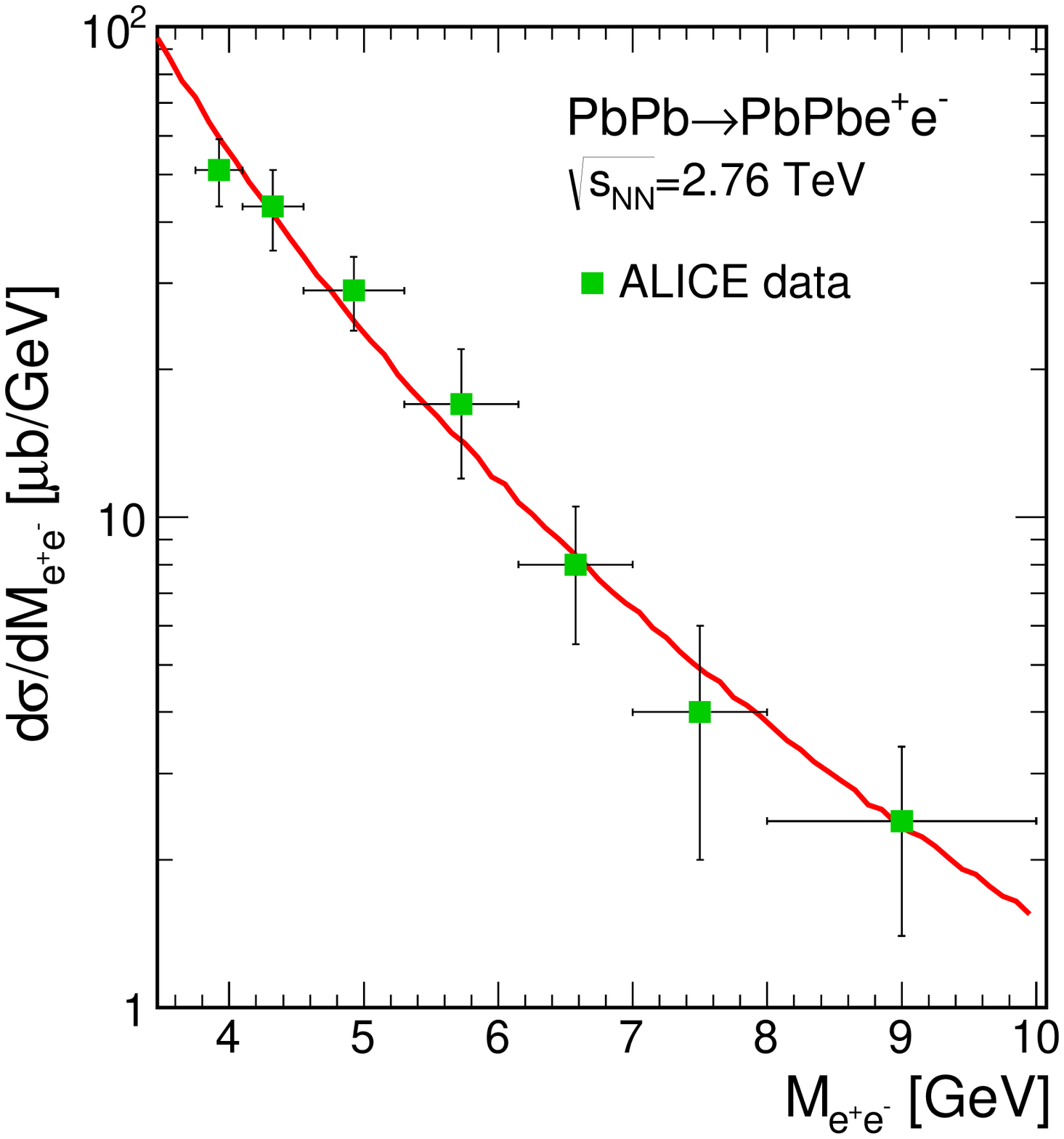}
\end{center}
\caption{Dielectron invariant mass
for the $Pb Pb \to Pb Pb e^+ e^-$ for the ALICE experimental cuts
\cite{ALICE_epem}.}
\label{fig:dsig_dMee}
\end{figure}

We start presentation of our results for dilepton production.
In Fig.\ref{fig:dsig_dMee} we present our results for dielectron
invariant mass together with ALICE experimental data \cite{ALICE_epem}.
A good agreement is achieved without free parameters.

\begin{figure}[!h]
\begin{center}
\includegraphics[scale=0.25]{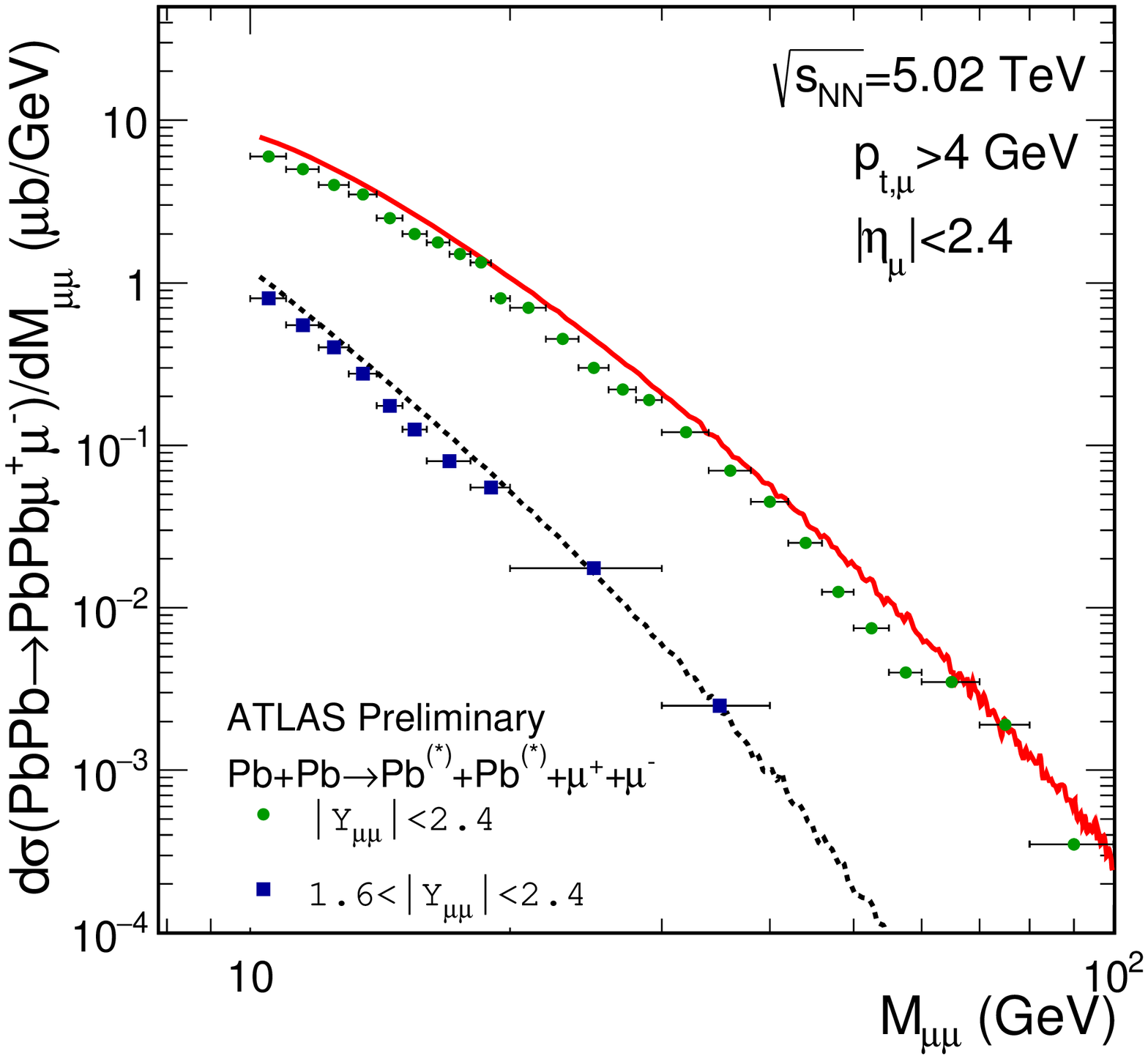}
\includegraphics[scale=0.25]{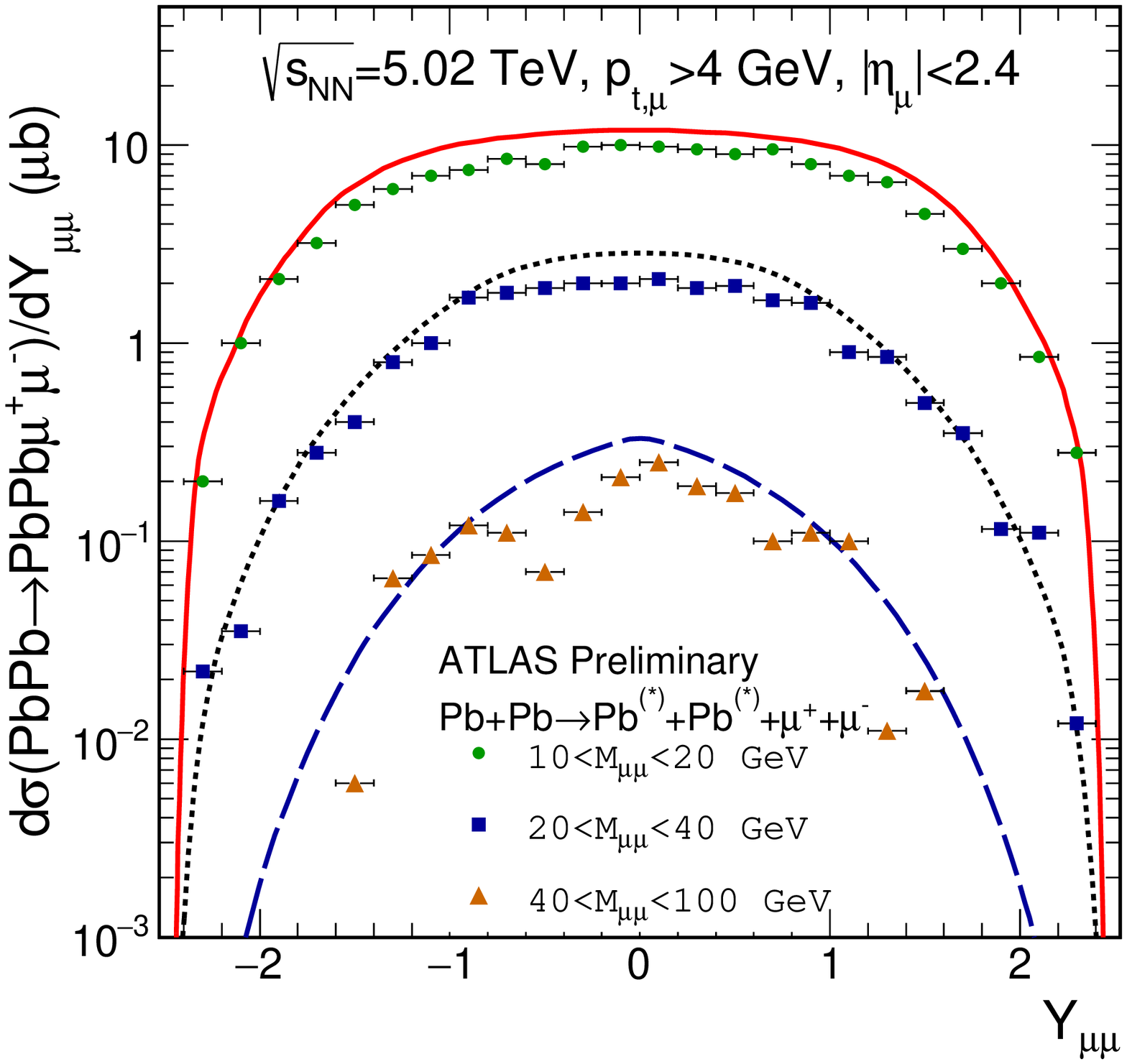}
\end{center}
\caption{Cross section for dimuon production in UPC of Pb+Pb
together with ATLAS experimental data \cite{ATLAS_mupmum}.}
\label{fig:dimuon}
\end{figure}

In Fig.\ref{fig:dimuon} we present our results together with 
the ATLAS data \cite{ATLAS_mupmum} for the $Pb+Pb \to Pb+Pb+\mu^++\mu^-$
reaction.
Experimental cuts were included in our calculations.
In contrast to the $Pb Pb \to Pb Pb e^+ e^-$ reaction the agreement
here is much worst.

\begin{figure}[h!]
\begin{center}
\includegraphics[scale=0.25]{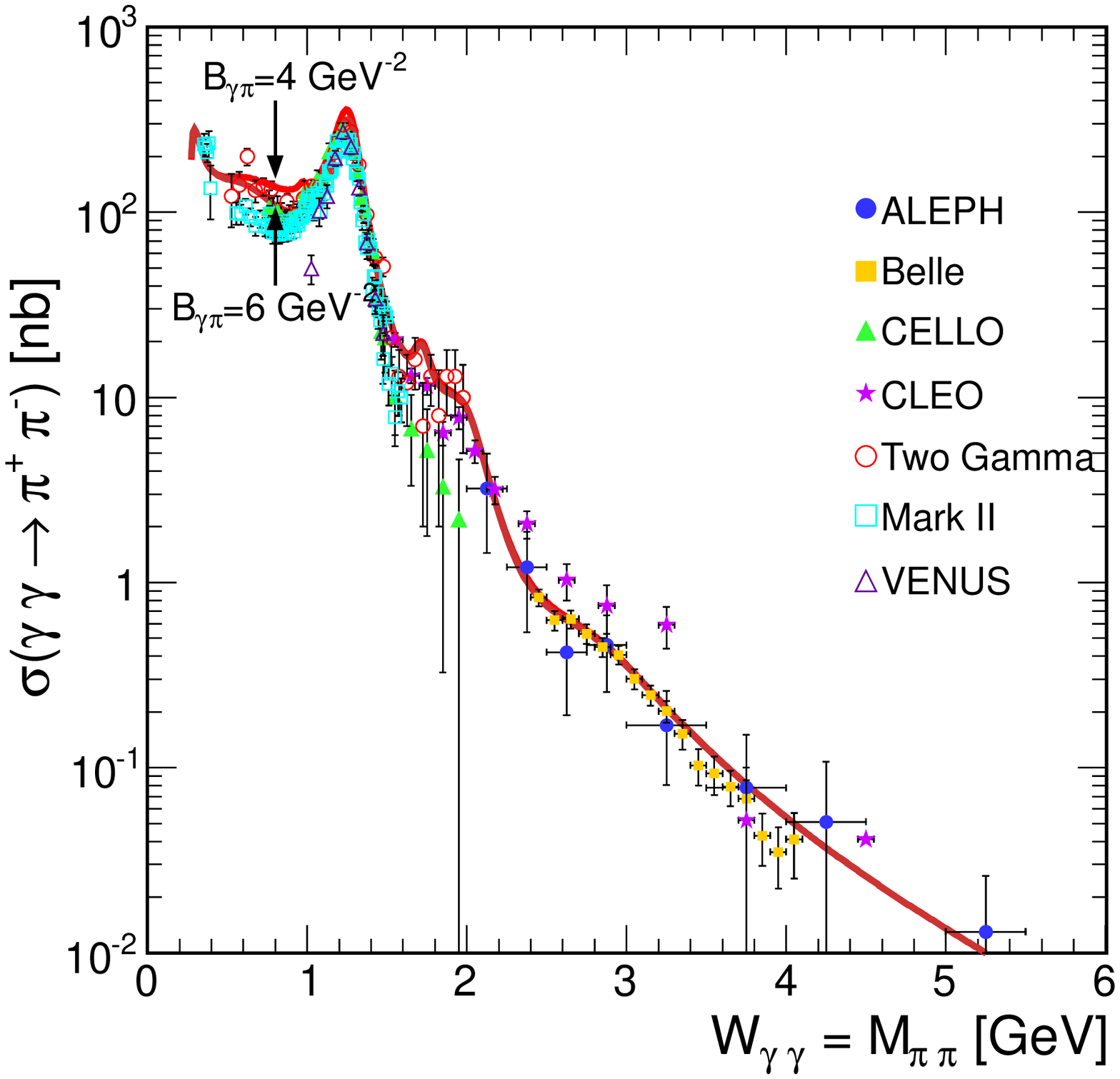}
\includegraphics[scale=0.25]{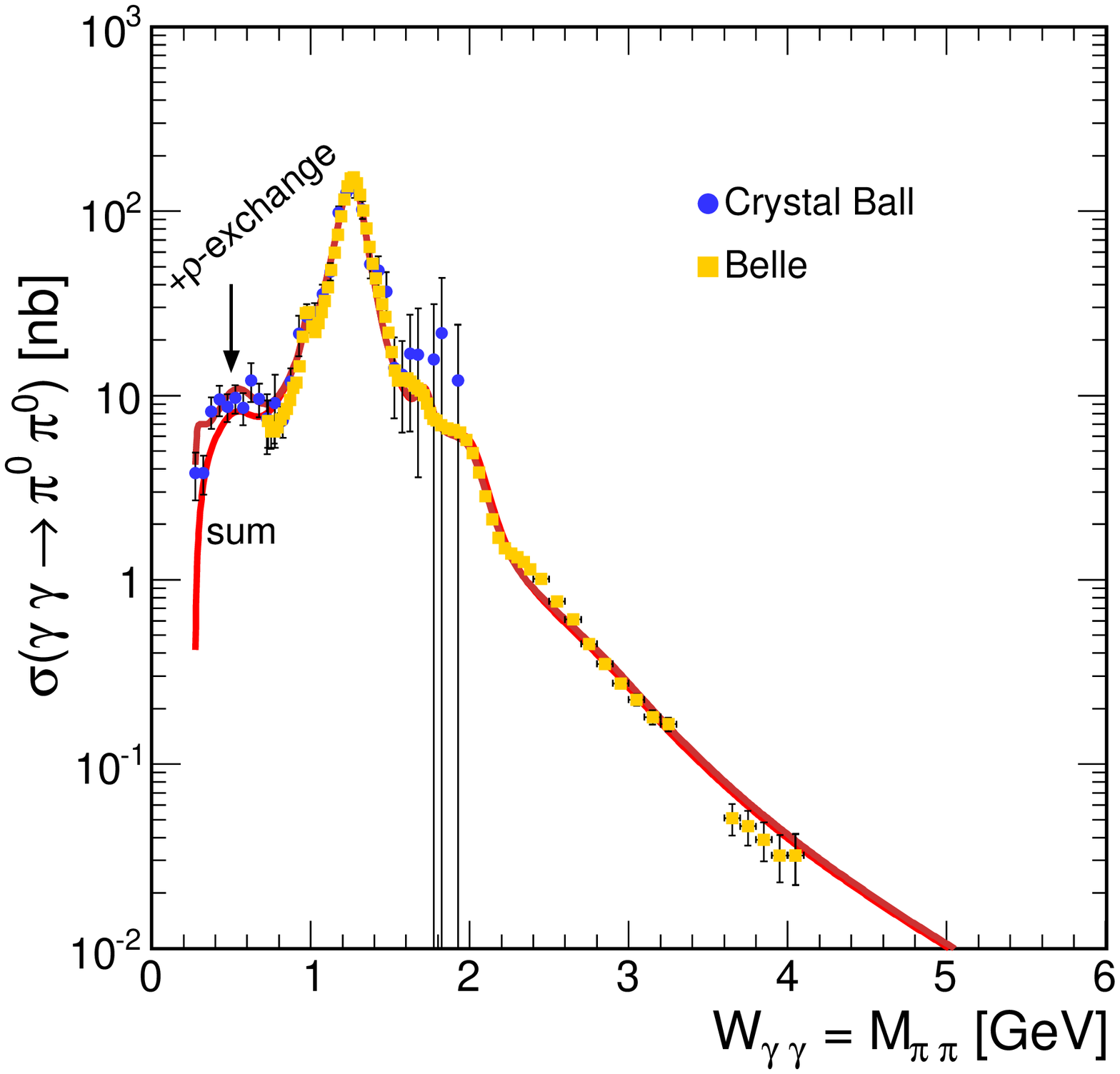}
\end{center}
\caption{Energy dependence of the elementary cross sections 
for $\gamma \gamma \to \pi \pi$ reactions.}
\label{fig:gamgam_pipi}
\end{figure}

Let us discuss now briefly production of charged pion pairs.
In Fig.\ref{fig:gamgam_pipi} we demonstrate how well our
multicomponent model \cite{KS2013} describes the elementary cross sections:
$\gamma \gamma \to \pi^+ \pi^-$ and $\gamma \gamma \to \pi^0 \pi^0$
measured in detail by different experiments \cite{KS2013}. 
These elementary cross sections can be used in calculation of the cross
section for nuclear processes $A A \to A A \pi \pi$.

\begin{figure}[!h]             
\begin{center}
\includegraphics[width=6cm]{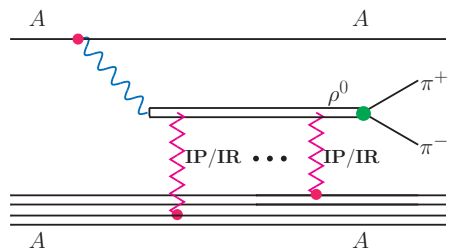}
\end{center}
\caption{Mechanism of coherent $\rho^0$ production.}
\label{fig:coherent_rho0}
\end{figure}

\begin{figure}[!h]
\begin{center}
\includegraphics[width=6cm]{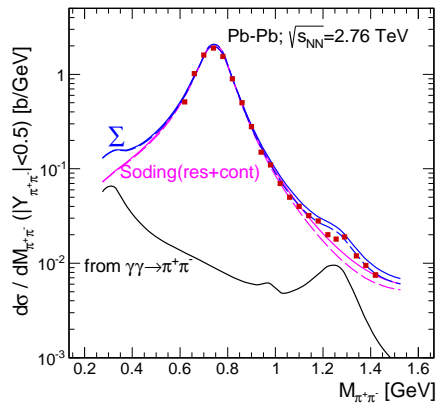}
\end{center}
\caption{Dipion invariant mass together with ALICE experimental data 
\cite{ALICE_pippim}. 
The $\rho^0 \to \pi^+ \pi^-$ and $\gamma \gamma \to \pi^+ \pi^-$ 
contributions are shown separately.}
\label{fig:dsig_dMpippim}
\end{figure}

There is a strong competition in the $\pi^+ \pi^-$ channel of
coherent $\rho^0$ meson production (see Fig.\ref{fig:coherent_rho0})
which decays into a $\pi^+ \pi^-$ pair.
The main mechanism is photon fluctuation into virtual $\rho^0$ meson
and its multiple rescattering in the collision partner.
In Fig.\ref{fig:dsig_dMpippim} we show invariant mass distribution
of the $\pi^+ \pi^-$ system. Both $\rho^0$ contribution with effective 
inclusion of the photoproduction continum, called sometimes S\"oding
mechanism, and the $\gamma \gamma$ mechanism were considered.
The $\gamma \gamma$ mechanism becomes sizeable in the region
of the $f_2(1270)$ dipion resonance and its presence improves
agreement with the ALICE experimental data \cite{ALICE_pippim}.

\begin{figure}[!h]
\begin{center}
\includegraphics[scale=0.25]{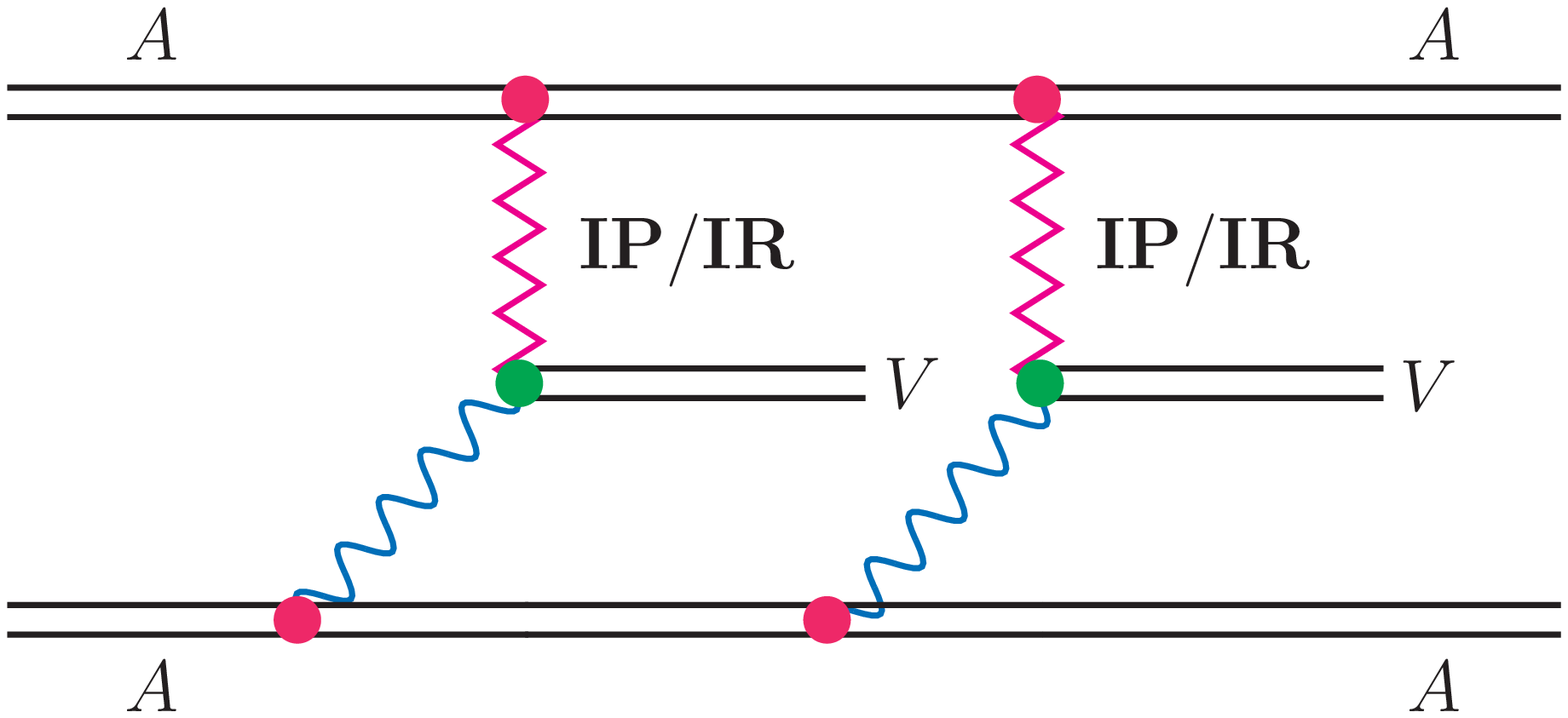}
\includegraphics[scale=0.25]{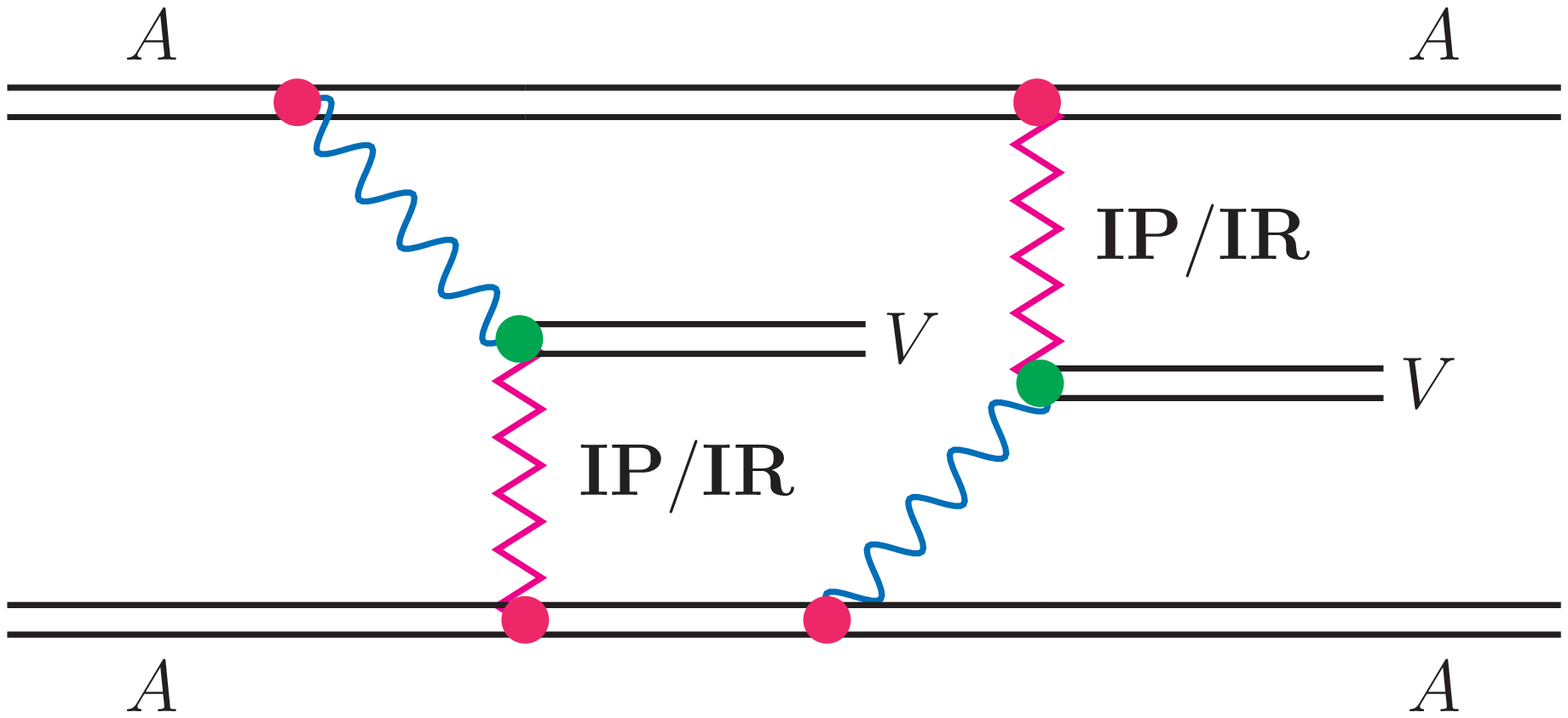}\\
\includegraphics[scale=0.25]{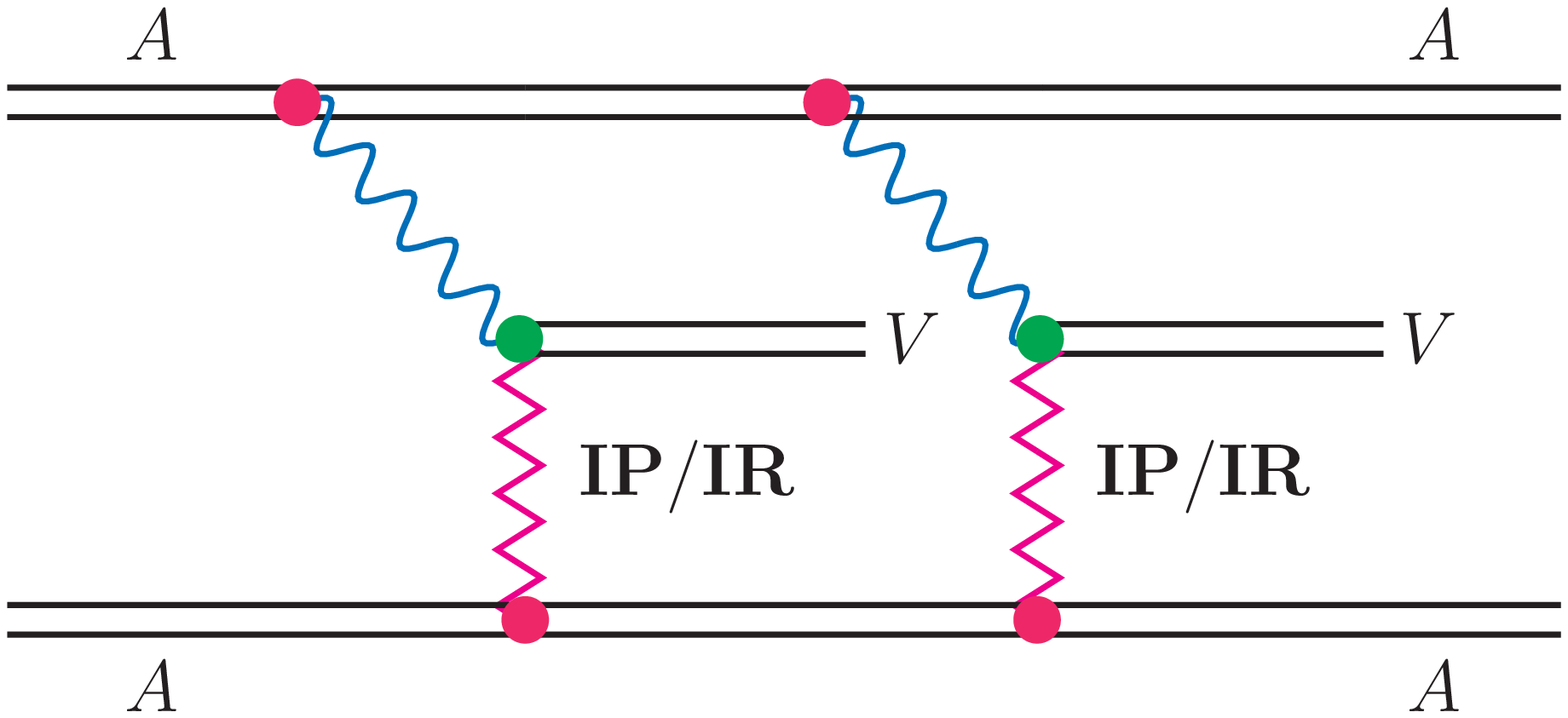}
\includegraphics[scale=0.25]{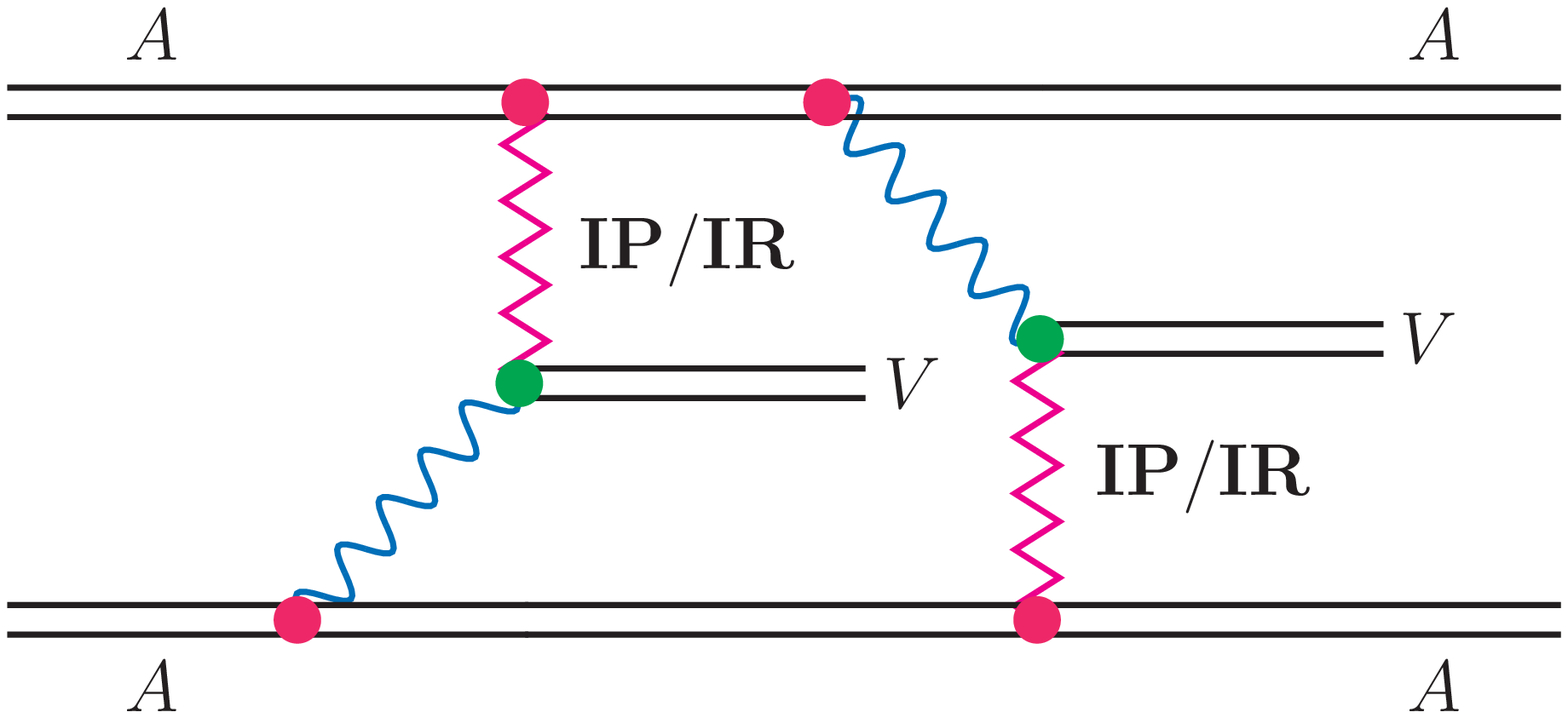}
\end{center}
\caption{The mechanisms of double $\rho^0$ production.}
\label{fig:rho0rho0_double_scattering}
\end{figure}

The cross section for coherent single $\rho^0$ production is very large.
Therefore one could consider also
double-scattering cross section for production of $\rho^0 \rho^0$ pairs.
The underlying mechanisms are sketched in 
Fig.\ref{fig:rho0rho0_double_scattering}. 

\begin{figure}
\begin{center}
\includegraphics[width=5cm]{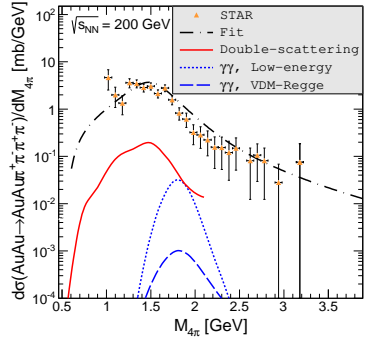}
\end{center}
\caption{Four-pion invariant mass distribution calculated by us together
with the STAR experimental data \cite{STAR_4pi}.
The contribution of the double scattering mechanism is shown by the
blue solid line. In addition we show contribution of single scattering
based on $\gamma \gamma \to \rho^0 \rho^0$ subprocess subdivided into
two subcontributions described in \cite{KSS2009}.}
\label{fig:dsig_dM4pi_STAR}
\end{figure}

In Fig.\ref{fig:dsig_dM4pi_STAR} we show distribution in four-pion invariant
mass for $\sqrt{s_{NN}}$ = 200 GeV together with STAR data 
\cite{STAR_4pi}.
We show the $\gamma \gamma \to \rho^0 \rho^0$
and double-scattering contributions. Clearly the double scattering
contribution is larger than the $\gamma \gamma$ one but insufficient 
to understand the STAR data \cite{STAR_4pi}. Is the disagreement 
due to coherent production of $\rho'$ or $\rho''$ mesons ? 
This is not clear in the moment and requires further studies in future.

\begin{figure}
\begin{center}
\includegraphics[width=5cm]{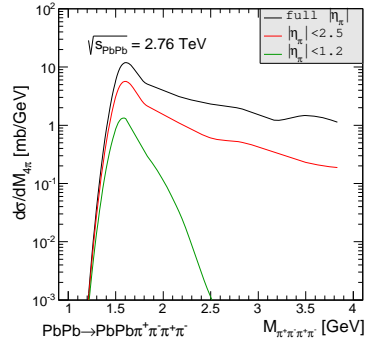}
\end{center}
\caption{Four-pion invariant mass at the LHC for different ranges
of pion pseudorapidity.}
\label{fig:dsig_dM4pi_LHC}
\end{figure}

In Fig.\ref{fig:dsig_dM4pi_LHC} we show our predictions
for four-pion invariant mass, including
only double-scattering mechanism for $\sqrt{s_{NN}}$ = 2.76 TeV.
The resulting distribution strongly depends on the range of rapidity.
A longer range is preferred when one wants to enhance the
double-scattering contribution.

\begin{figure}[!h]
\begin{center}
\includegraphics[width=4.5cm]{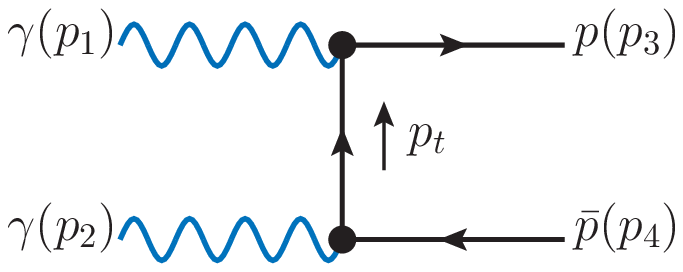}
\includegraphics[width=4.5cm]{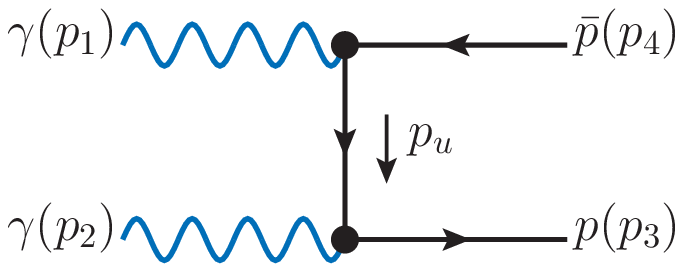}
\end{center}
\caption{Elementary processes $\gamma \gamma \to p \bar p$
responsible for production of $p \bar p$ pairs in UPC of heavy ions.}
\label{fig:gamgam_ppbar}
\end{figure}

\begin{figure}
\begin{center}
\includegraphics[scale=0.25]{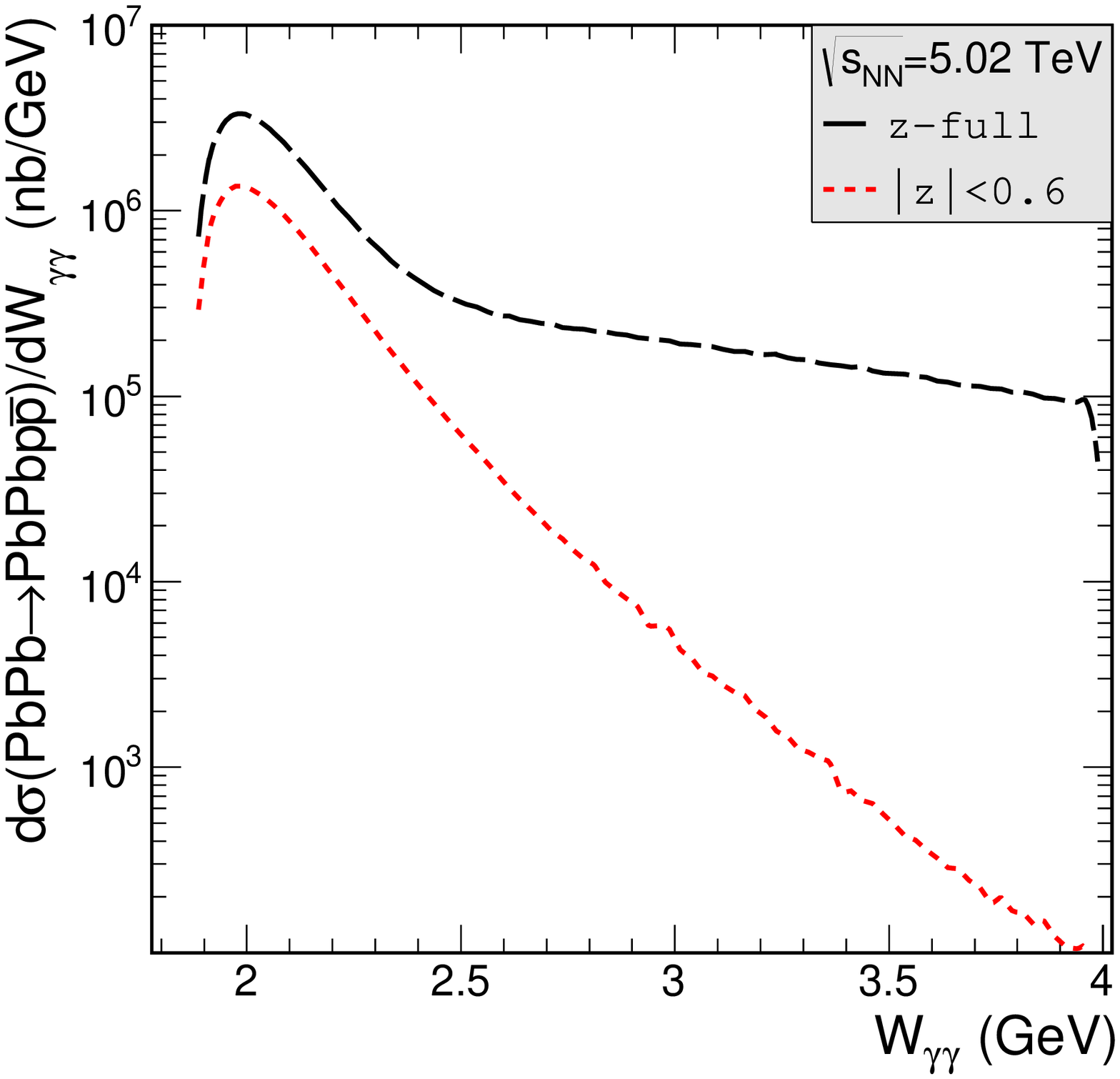}
\includegraphics[scale=0.25]{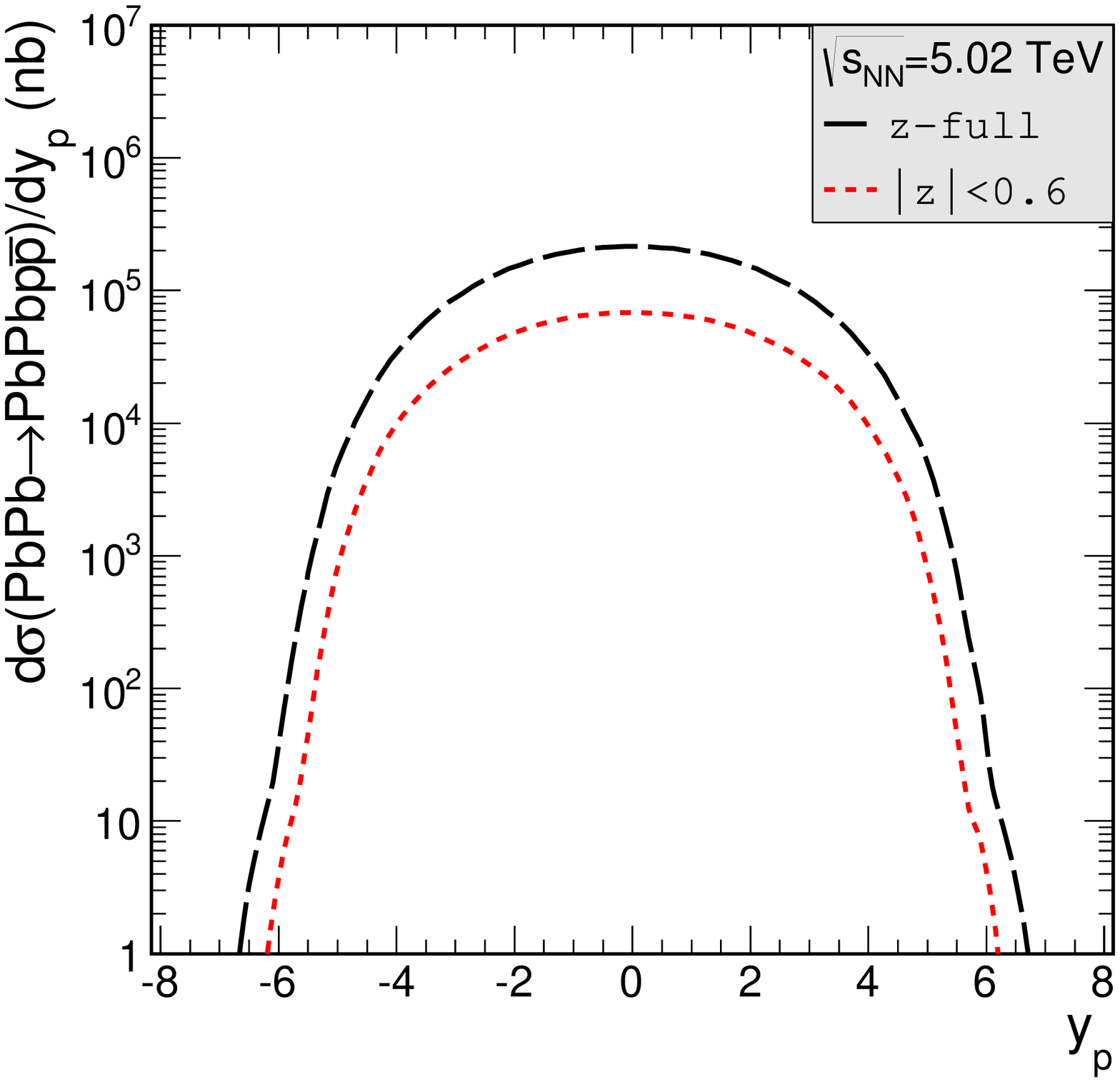}
\end{center}
\caption{Examples of differential cross sections for the 
$Pb Pb \to Pb Pb p \bar p$ reaction.}
\label{fig:AA_AAppbar}
\end{figure}

Another interesting process is $A A \to A A p \bar p$.
The continuum subprocess is shown for example in
Fig.\ref{fig:gamgam_ppbar}.
In our studies we included also some resonances \cite{KLNS2017}.
In Fig.\ref{fig:AA_AAppbar} we show our predictions
for $M_{p \bar p}$ and rapidity distributions for 
$Pb Pb \to Pb Pb p \bar p$ process at $\sqrt{s_{NN}}$ = 5.02 TeV.
Predicted cross sections for $Pb Pb \to Pb Pb p \bar p$ 
for different experimental cuts are given in Tab.1.

{\scriptsize{\begin{table}[!h]
    \begin{tabular}{|c|l|c|}
    \hline
    Experiment & Cuts & $\sigma$ [$\mu$b] \\ \hline
    ALICE 	& $p_{t,p}>$ 0.2 GeV,
    $|y_p| <$ 0.9 & 100  \\
    ATLAS	& $p_{t,p}>0.5$ GeV,
    $|y_p| <$ 2.5 & 160  \\
    CMS		& $p_{t,p}>0.2$ GeV,
    $|y_p| <$ 2.5 & 500  \\
    LHCb	& $p_{t,p}>0.2$ GeV,
    2 $< y_p <$ 4.5 & 104  \\
    \hline
    \end{tabular}
    \end{table}}
            		}

\begin{figure}[!h]
\begin{center}
\includegraphics[width=8cm]{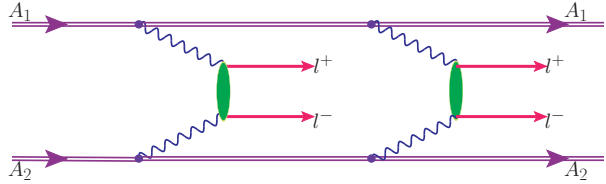}
\end{center}
\caption{Double scattering production mechanism of two lepton pairs.}
\label{fig:AA_AA4leptons}
\end{figure}

\begin{figure}
\begin{center}
\includegraphics[scale=0.25]{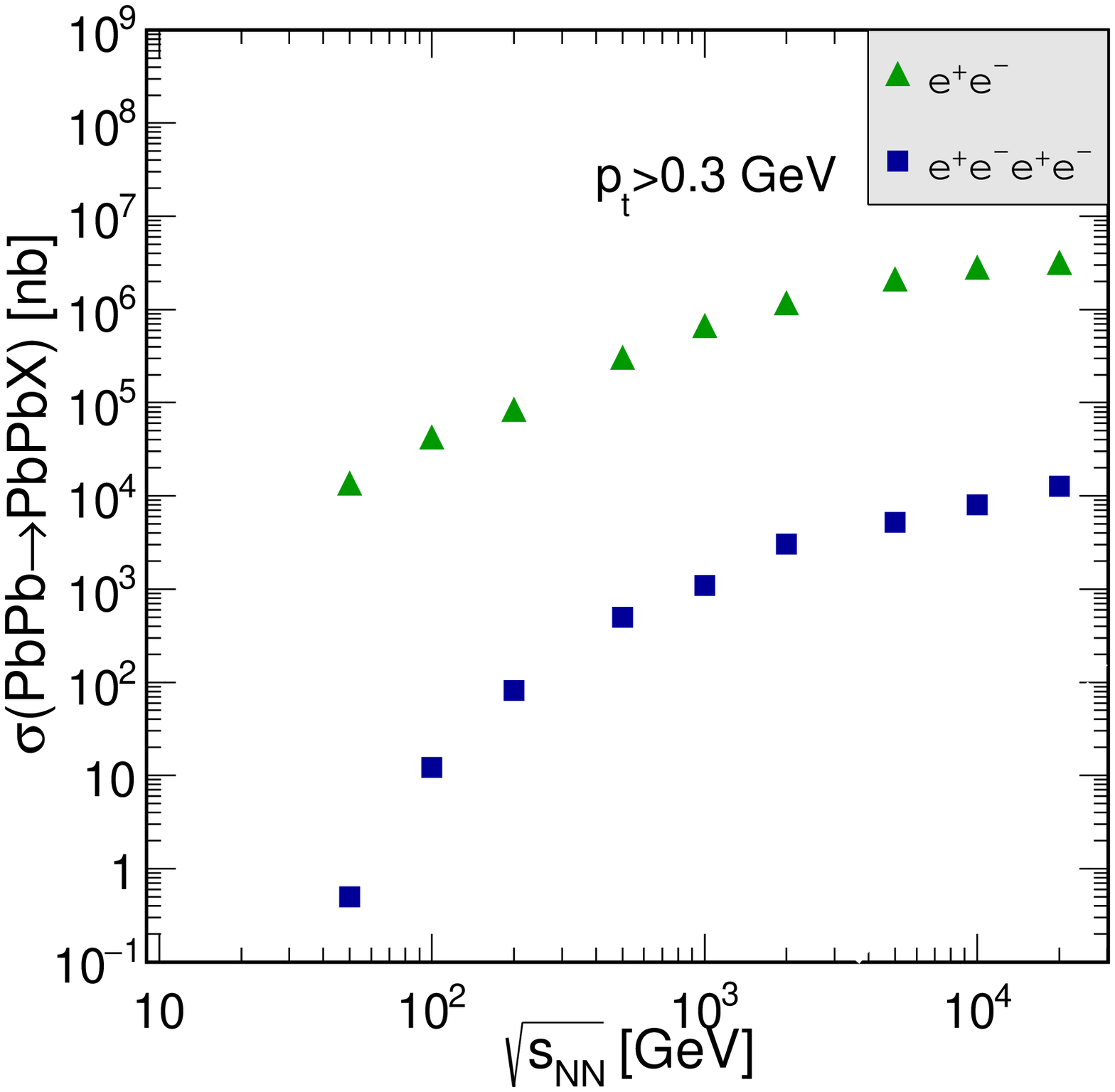}
\includegraphics[scale=0.25]{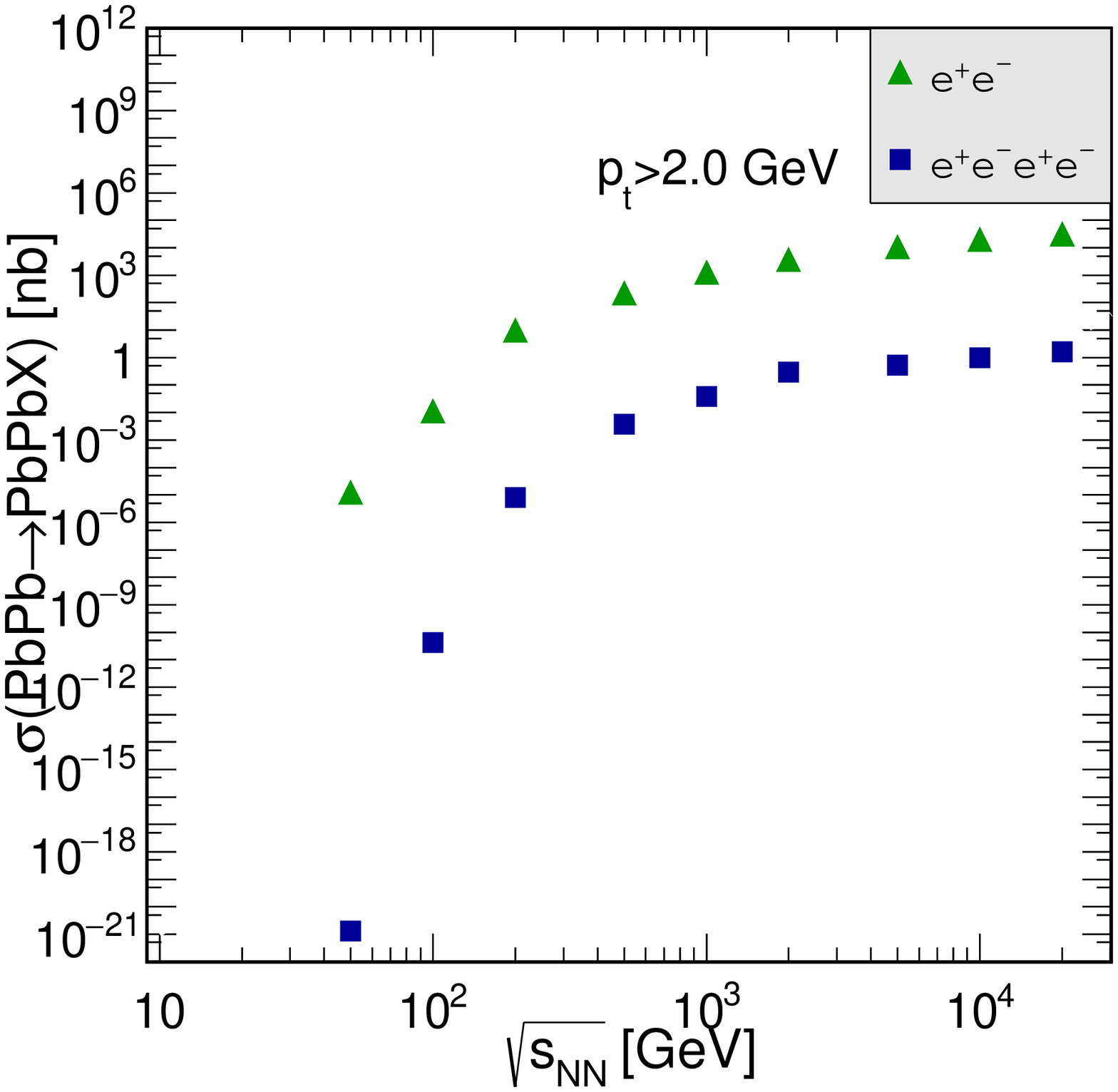}
\end{center}
\caption{Phase-space integrated cross section
for $e^+ e^- e^+ e^-$ and $e^+ e^-$ production for two different cuts
on lepton transverse momenta.}
\label{fig:4eresults}
\end{figure}

Double scattering UPC are possible also for production of two
lepton pairs as shown in Fig.\ref{fig:AA_AA4leptons}.
The cross section integrated over phase space 
is shown in Fig.\ref{fig:4eresults} for two different cuts on 
lepton transverse momenta (the same for each lepton).

The number of counts for integrated luminosity $L_{int}$ = 1 nb$^{-1}$
is given in Tab.2. The table shows that some measurements
of four leptons are possible. Certainly such a test of our predictions
of double scattering mechanism would be new and valueable.

               \begin{table}[!h]
               	\begin{tabular}{|c|r||c|r|}
               		\hline
               		\multicolumn{2}{|c||}{($4\mu$), $\sqrt{s_{NN}}=5.02$ TeV} &
               		\multicolumn{2}{|c||}{($4e$), $\sqrt{s_{NN}}=5.5$ TeV} \\
               		experimental cuts       & N & experimental cuts       & N \\         
               		\hline
               		$|y_i| <$ 2.5, $p_t >$ 0.5 GeV   	&  815 & $|y_i| <$ 2.5, $p_t >$ 0.5 GeV & 235 \\
               		$|y_i| <$ 2.5, $p_t >$ 1.0 GeV  	&   53 & $|y_i| <$ 2.5, $p_t >$ 1.0 GeV & 10 \\
               		$|y_i| <$ 0.9, $p_t >$ 0.5 GeV  	&   31 & $|y_i| <$ 1.0, $p_t >$ 0.2 GeV & 649 \\
               		$|y_i| <$ 0.9, $p_t >$ 1.0 GeV   	&    2 & $|y_i| <$ 1.0, $p_t >$ 1.0 GeV & 1 \\
               		$|y_i| <$ 2.4, $p_t >$ 4.0 GeV  &   $\ll$1 & \multicolumn{2}{|c|}{ } \\
               		\hline
               	\end{tabular}
               \end{table}

\begin{figure}
\begin{center}
\includegraphics[scale=0.35]{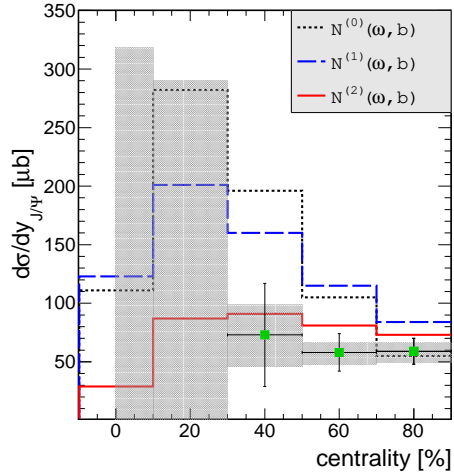}
\end{center}
\caption{Dependence of the cross section for creation of the $J/\psi$
  meson as a function of meson centrality together with the ALICE
experimental data \cite{ALICE_jpsi}.}
\label{fig:jpsi_notonlyUPC}
\end{figure}

Many of the processes discussed here survive also to the situation
when the nuclei collide and when centrality of the collision can be
determined. A first example was discussed for photoproduction
of $J/\psi$ quarkonium in \cite{KS2016}.
In Fig.\ref{fig:jpsi_notonlyUPC} we show the cross section for different
bins of centrality. Rather good description of the data was achieved
by imposing special conditions on photon fluxes \cite{KS2016}.

Another example is the $AA \to e^+ e^-$ peripheral and semicentral
nucleus-nucleus collisions for small dilepton
transverse momenta discussed very recently \cite{KRSS2018}.
The photoproduction mechanism is particularly important for small
dielectron transverse momenta and not too small energies where
it competes with thermal dielectron production.

\vspace{1cm}

This work was supported by the National Science Centre, Poland (NCN),
grant number 2014/15/B/ST2/02528.





\begin{thebibliography}{99}

\bibitem{reviews}
V.M. Budnev, I.F. Ginzburg, G.V. Meledin and V.G. Serbo,
Phys. Rep. {\bf 15} (1975) 4;
C.A.~Bertulani and G. Baur, 
Phys. Rep. {\bf 163} (1988) 29;
G. Baur, K. Hencken, D. Trautmann, S.~Sadovsky, 
and Y. Kharlov, Phys. Rep. {\bf 364} (2002) 359;
A.J. Baltz, G. Baur, D.~d'Enterria et al., 
Phys. Rep. {\bf 458} (2008) 1.

\bibitem{KS2010} 
M. K{\l}usek-Gawenda and A. Szczurek,
``Exclusive muon-pair production in utrarelativistic heavy-ion
collisions - realistic nucleus charge form factor and differential
distributions'',
Phys. Rev. {\bf C82} (2010) 014904.

\bibitem{KSMS2011} 
M. K{\l}usek-Gawenda, A. Szczurek, M. Machado and V. Serbo,
``Double -- photon exclusive processes with heavy quark -- heavy
antiquark pairs in high-energy Pb-Pb collisions at LHC'',
Phys. Rev. {\bf C83} (2011) 024903.

\bibitem{KSS2009}
M. K{\l}usek, W. Sch\"afer and A. Szczurek,
"Exlusive production of $\rho^0 \rho^0$ pairs
in $\gamma \gamma$ collisions at RHIC",
Phys. Lett. {\bf B674} (2009) 92.
 
\bibitem{KS2013} 
M. K{\l}usek-Gawenda and A. Szczurek,
``$\pi^+ \pi^-$ and $\pi^0 \pi^0$ pair production in photon-photon
and in ultraperipheral ultrarelativistic heavy ion collisions'',
Phys. Rev. {\bf C87} (2013) 054908.

\bibitem{KS2011} 
M. K{\l}usek-Gawenda and A. Szczurek,
``Exclusive production of large invariant mass pion pairs in
ultrarelativistic heavy ion collisions'',
Phys. Lett. {\bf B700} (2011) 322.

\bibitem{BCKSS2013}  
S. Baranov, A. Cisek, M. K{\l}usek-Gawenda, W. Sch\"afer
and A. Szczurek,\\
"The $\gamma \gamma \to J/\psi J/\psi$ reaction and
the $J/\psi J/\psi$ pair production  
in exclusive ultraperipheral ultrarelativistic heavy ion collisions",
Eur. Phys. J. {\bf C73} (2013) 2335. 

\bibitem{KS2014} 
M. K{\l}usek-Gawenda and A. Szczurek,
"Double-scattering mechanism in the exclusive $A A \to A A \rho^0 \rho^0$ 
reaction in ultrarelativistic collisions",
Phys. Rev. {\bf C89} (2014) 024912.

\item M. K{\l}usek-Gawenda, P. Lebiedowicz, and A. Szczurek, 
``Light-by-light scattering in ultraperipheral Pb-Pb collisions at
energies available at the CERN Large Hadron Collider'',
Phys. Rev. {\bf C93} (2016) 044907.

\bibitem{KSS2016} 
M. K{\l}usek-Gawenda, W. Sch\"afer and A. Szczurek,
``Two-gluon exchange contribution to elastic $\gamma \gamma \to \gamma \gamma$ 
scattering and production of two-photons in ultraperipheral
ultrarelativistic heavy ion and proton-proton collisions'',
Phys. Lett. {\bf B761} (2016) 399.

\bibitem{KLNS2017} 
M. K{\l}usek-Gawenda, P. Lebiedowicz, O. Nachtmann and A. Szczurek,
``From the $\gamma \gamma \rightarrow p \overline{p}$ reaction
to the production of $p \overline{p}$ pairs in ultrarelativistic
heavy-ion collisions at the LHC",
Phys. Rev. {\bf D96} (2017) 094029.

\bibitem{KS2016}
M. K{\l}usek-Gawenda and A. Szczurek,
``Double scattering production of two positron-electron pairs
in ultraperipheral heavy-ion collisions'',
Phys. Lett. {\bf B763} (2016) 416.

\bibitem{HKS2018} 
A. Hameren, M. K{\l}usek-Gawenda and A. Szczurek,
``Single- and double-scattering production of four muons 
in ultraperipheral $PbPb$ collisions 
at the Large Hadron Collider'',
Phys. Lett. {\bf B776} (2018) 84.

\bibitem{KS2016_jpsi_semicentral} 
M. K{\l}usek-Gawenda and A. Szczurek,
`` Photoproduction of $J/\psi$  mesons in peripheral and semicentral heavy
   ion collisions'',
Phys. Rev. {\bf C93} (2016) 044912.

\bibitem{KCSS2014} 
M. K{\l}usek-Gawenda, M. Ciema{\l}a, W. Sch\"afer and A. Szczurek,
``Electromagnetic excitation of nuclei and neutron evaporation
in ultrarelativistic ultraperipheral heavy ion collisions'',
Phys. Rev. {\bf C89} (2014) 054907.

\bibitem{KRSS2018}
M. K{\l}usek-Gawenda, R. Rapp, W. Sch\"afer and A. Szczurek,
arXiv:1809.07049.

\bibitem{Klusek_talk}
M. K{\l}usek-Gawenda, a talk at this conference.

\bibitem{ALICE_epem}
E. Abbas et al. (ALICE collaboration), Eur.Phys.J. {\bf C73} (2013) 2617.

\bibitem{ATLAS_mupmum}
ATLAS collaboration, ATLAS-CONF-2016-025.

\bibitem{ALICE_pippim}
J. Adam et al. (ALICE collaboration), JHEP{\bf 09} (2015) 095.

\bibitem{STAR_4pi}
B.I. Abelev et al. (STAR collaboration), Phys.Rev. {\bf C81} (2010) 044901.

\bibitem{ALICE_jpsi}
J. Adam et al. (ALICE collaboration), Phys.Rev.Lett.{\bf 116} (2016) 222301.


\end{thebibliography}
\end{document}